\documentclass[aps,prl,superscriptaddress,twocolumn,showpacs,amsmath,amssymb,floatfix]{revtex4}% Physical Review Letters

\usepackage{graphicx}% Include figure files
\usepackage{dcolumn}% Align table columns on decimal point
\usepackage{bm}% bold math

\def\casmu {\Xi^0 \rightarrow \Sigma^+\mu^-\bar{\nu}_{\mu}}
\def\casbet{\Xi^0 \rightarrow \Sigma^+e^-\bar{\nu}_e}
\def\caslpi{\Xi^0 \rightarrow \Lambda\pi^0}
\def\ktpi  {K_L \rightarrow \pi^+\pi^-\pi^0}
\def\caslg{\Xi^0 \rightarrow \Lambda\gamma}
\def\cassg{\Xi^0 \rightarrow \Sigma^0\gamma}
\def\lamppi{\Lambda \rightarrow p\pi^-}
\def\lammu{\Lambda \rightarrow p\mu^-\bar{\nu}_{\mu}}

\def\kmu4{K_L \rightarrow \pi^0\pi^+\mu^-\bar{\nu}_{\mu}}
\def\siglam{\Sigma^0 \rightarrow \Lambda\gamma}

\def\UAz{University of Arizona, Tucson, Arizona 85721}
\def\UCLA{University of California at Los Angeles, Los Angeles, California 90095}
\def\Campinas{Universidade Estadual de Campinas, Campinas, Brazil 13083-970}
\def\EFI{The Enrico Fermi Institute, The University of Chicago, Chicago, Illinois 60637}
\def\UB{University of Colorado, Boulder, Colorado 80309}
\def\ELM{Elmhurst College, Elmhurst, Illinois 60126}
\def\FNAL{Fermi National Accelerator Laboratory, Batavia, Illinois 60510}
\def\Osaka{Osaka University, Toyonaka, Osaka 560-0043 Japan}
\def\Rice{Rice University, Houston, Texas 77005}
\def\SaoPaulo{Universidade de S\~ao Paulo, S\~ao Paulo, Brazil 05315-970}
\def\UVa{The Department of Physics and Institute of Nuclear and
         Particle Physics, University of Virginia, Charlottesville, Virginia 22901}
\def\UW{University of Wisconsin, Madison, Wisconsin 53706}

\begin{document}

\title{Observation of the Decay $\mathbf \casmu$}

\affiliation{\UAz}
\affiliation{\UCLA}
\affiliation{\Campinas}
\affiliation{\EFI}
\affiliation{\UB}
\affiliation{\ELM}
\affiliation{\FNAL}
\affiliation{\Osaka}
\affiliation{\Rice}
\affiliation{\SaoPaulo}
\affiliation{\UVa}
\affiliation{\UW}

\author{E.~Abouzaid}      \affiliation{\EFI}
\author{T.~Alexopoulos}   \affiliation{\UW}
\author{M.~Arenton}       \affiliation{\UVa}
\author{R.~F.~Barbosa}    \affiliation{\SaoPaulo}
\author{A.~R.~Barker}     \altaffiliation[Deceased]{}\affiliation{\UB}
\author{L.~Bellantoni}    \affiliation{\FNAL}
\author{A.~Bellavance}    \affiliation{\Rice}
\author{E.~Blucher}       \affiliation{\EFI}
\author{G.~J.~Bock}       \affiliation{\FNAL}
\author{E.~Cheu}          \affiliation{\UAz}
\author{R.~Coleman}       \affiliation{\FNAL}
\author{M.~D.~Corcoran}   \affiliation{\Rice}
\author{B.~Cox}           \affiliation{\UVa}
\author{A.~R.~Erwin}      \affiliation{\UW}
\author{C.~O.~Escobar}    \affiliation{\Campinas}
\author{A.~Glazov}        \affiliation{\EFI}
\author{A.~Golossanov}    \affiliation{\UVa}
\author{R.~A.~Gomes}      \altaffiliation[To whom correspondence should be addressed.\\
Electronic address: {\bf ragomes@ifi.unicamp.br}]{} \affiliation{\Campinas}
\author{P.~Gouffon}       \affiliation{\SaoPaulo}
\author{K.~Hanagaki}      \affiliation{\Osaka}
\author{Y.~B.~Hsiung}     \affiliation{\FNAL}
\author{H.~Huang}         \affiliation{\UB}
\author{D.~A.~Jensen}     \affiliation{\FNAL}
\author{R.~Kessler}       \affiliation{\EFI}
\author{K.~Kotera}        \affiliation{\Osaka}
\author{J.~LaDue}         \affiliation{\UB}
\author{A.~Ledovskoy}     \affiliation{\UVa}
\author{P.~L.~McBride}    \affiliation{\FNAL}
\author{E.~Monnier}       \altaffiliation[Permanent address: ]{C.P.P. Marseille/C.N.R.S., France} \affiliation{\EFI}
\author{K.~S.~Nelson}     \affiliation{\UVa}
\author{H.~Nguyen}        \affiliation{\FNAL}
\author{R.~Niclasen}      \affiliation{\UB}
\author{H.~Ping}          \affiliation{\UW}
\author{V.~Prasad}        \affiliation{\EFI}
\author{X.~R.~Qi}         \affiliation{\FNAL}
\author{E.~J.~Ramberg}    \affiliation{\FNAL}
\author{R.~E.~Ray}        \affiliation{\FNAL}
\author{M.~Ronquest}      \affiliation{\UVa}
\author{E.~Santos}        \affiliation{\SaoPaulo}
\author{J.~Shields}       \affiliation{\UVa}
\author{W.~Slater}        \affiliation{\UCLA}
\author{D.~Smith}         \affiliation{\UVa}
\author{N.~Solomey}       \affiliation{\EFI}
\author{E.~C.~Swallow}    \affiliation{\EFI}\affiliation{\ELM}
\author{P.~A.~Toale}      \affiliation{\UB}
\author{R.~Tschirhart}    \affiliation{\FNAL}
\author{C.~Velissaris}    \affiliation{\UW}
\author{Y.~W.~Wah}        \affiliation{\EFI}
\author{J.~Wang}          \affiliation{\UAz}
\author{H.~B.~White}      \affiliation{\FNAL}
\author{J.~Whitmore}      \affiliation{\FNAL}
\author{M.~Wilking}       \affiliation{\UB}
\author{B.~Winstein}      \affiliation{\EFI}
\author{R.~Winston}       \affiliation{\EFI}
\author{E.~T.~Worcester}  \affiliation{\EFI}
\author{M.~Worcester}     \affiliation{\EFI}
\author{T.~Yamanaka}      \affiliation{\Osaka}
\author{E.~D.~Zimmerman}  \affiliation{\UB}
\author{R.~F.~Zukanovich$^{10}$\\{\bf The KTeV Collaboration}}
%\affiliation{\SaoPaulo}

\date{\today}

\begin{abstract}
The $\Xi^0$ muon semi-leptonic decay has been observed for the first
time with nine identified events using the KTeV beam line and detector
at Fermilab. The decay is normalized to the $\Xi^0$ beta decay mode
and yields a value for the ratio of decay rates
$\Gamma(\casmu)/\Gamma(\casbet)$ of
$(1.8^{+0.7}_{-0.5}(stat.)\pm0.2(syst.))\times 10^{-2}$. This is in
agreement with the SU(3) flavor symmetric quark model.
\end{abstract}


\maketitle

We report the first observation of the $\Xi^0$ muon semi-leptonic
decay, $\casmu$, and measurement of its decay rate normalized to the
topologically identical $\Xi^0$ beta decay, $\casbet$. This
measurement was performed at the KTeV experiment at Fermilab, using
methods similar to the first observation of the $\Xi^0$ beta decay
\cite{affolder}. The present observation is a new contribution to the
study of hyperon semi-leptonic decays, whose study elucidates the
structure of hadrons \cite{csw}.

The flavor symmetric quark model with the Cabbibo-Koboyashi-Maskawa
\cite{ckm} matrix elements and form factors obtained from Baryon
semi-leptonic decays can be used to predict \cite{garcia} the decay
rate of $\Xi^0 \rightarrow \Sigma^+ \ell^- \overline{\nu}_{\ell}$,
where $\ell$ = $e$ or $\mu$.  Lepton flavor symmetry (traditionally
called $e\!\!-\!\!\mu$ universality in this context) requires that the
$e$ and $\mu$ mode decay rates differ only through the differing
charged lepton mass values which appear in phase space factors,
radiative corrections and terms higher order in $\Delta M/M$, where
$\Delta M$ is the difference between initial and final baryon
masses. A calculation using the form factors described in
Ref.~\cite{garcia} and the latest value for the $\Xi^0$ mass from
Ref.~\cite{pdg} yields $\Gamma(\casmu)/\Gamma(\casbet) = 0.9\times
10^{-2}$.  Therefore the observation of the $\Xi^0$ muonic decay mode
and its ratio to the electron mode serves as a test of the Standard
Model description of these decays and the assumption of lepton
universality.

The observation reported here is based on the 1999 data set collected
during the E799-II (rare decay) configuration of KTeV \cite{arisaka}.
The experiment, while mostly known as a high precision investigation
of CP violation with an extensive rare kaon decay program, also
afforded the opportunity of studying neutral hyperons ($\Lambda$ and
$\Xi^0$). The KTeV beam line and detector used for these hyperon
studies have been extensively described in Ref.~\cite{affolder},
therefore only the main components are recalled. An intense
800~GeV/$c$ proton beam from the Tevatron was directed onto a BeO
target at a vertical angle of 4.8~mrad.  Photons were converted by a
lead absorber 20 m from the target and charged particles were swept
out of the beam by a series of dipole magnets. Collimators defined two
secondary neutral beams that entered a 65~m long vacuum tank, which
determined the decay region beginning at 94~m from the target.  The
integrated magnetic field from the sweeping magnets in the beam line
delivered $\Xi^0$ hyperons polarized (with about 10\% polarization) in
the positive or negative vertical direction.  Reversing the polarity
of a magnet called the {\it spin rotator dipole} regularly gave a net
polarization of zero for the data discussed here.  There were about $3
\times 10^8$ $\Xi^0$ and $3.5 \times 10^{11}$ $K_L$ decays in the
decay region. The momentum of the $\Xi^0$ was peaked at 290~GeV/$c$.

%\begin{figure}[t]
%\includegraphics[width=8cm]{fig01}
%\caption{\label{fig:detector} The KTeV detector apparatus as
%  configured for this measurement. The transition radiation detectors
%  (TRD and Beamline TRD) were not used in this analysis.}
%\end{figure}

The charged particle spectrometer consisted of two pairs of drift
chambers with a dipole magnet in between providing a transverse
momentum kick of $\pm$150 MeV/$c$. The electromagnetic calorimeter,
having energy resolution better than 1\% and position resolution of
$\sim$1~mm, was made of 3100 independent cesium iodide (CsI)
crystals. Two holes in the calorimeter allowed the beams to pass
through and impact on a beam calorimeter located further downstream.
In addition, various veto elements (Ring Vetos, Spectrometer Anti,
Collar-Anti and Back-Anti) were used to detect particles escaping the
fiducial volume of the detector. %(Figure~\ref{fig:detector}).

The muon identification system was designed to stop charged pions while
permitting muons to pass through.  It was composed of a 10~cm thick
lead wall followed by a series of three iron walls of 5~m total thickness. 
A large hole that enveloped both beams was in both the lead wall and first 
iron wall of 1~m thickness. Two scintillators in this hole were used to 
detect charged particles (e.g. the proton from $\Lambda\rightarrow p\pi^-$)
in either beam. Charged pion showers outside the beams were vetoed at the 
trigger level by the Hadron-Anti hodoscope behind the lead wall. 
The second iron wall was 3~m thick and had no hole. Immediately behind 
this wall was a muon hodoscope $(\mu 2)$ of slightly overlapping scintillator 
paddles. Downstream of this was an additional meter of
iron and two planes of muon hodoscopes ($\mu$3 planes), one horizontal
and another vertical.

In the reconstruction analysis, the final detectable state of $\casmu$
is a proton, a muon and two photons, $[p\mu^-\gamma\gamma]$,
considering the subsequent $\Sigma^+ \rightarrow p\pi^0$ and $\pi^0
\rightarrow \gamma\gamma$ decays. We do not detect the
neutrino. The only difference in the normalizing mode final state is
an electron instead of a muon, $[pe^-\gamma\gamma]$.  Therefore, it
was possible to include both $\Xi^0$ muon and beta decay modes in the
same trigger sample, allowing to cancel possible biases in event
reconstruction, since they have similar secondary decays.

The trigger system of KTeV used three levels to select the events to
keep on tape. The first and second levels used logical combinations of
signals from the electronic hardware, while the third level carried out a
fast online reconstruction allowing event selection based on physics
criteria.  The trigger required at least two neutral clusters ({\em i.e.}
not associated with charged particle tracks) of energy deposited in the
electromagnetic calorimeter, presumably by the two photons of the $\pi^0$
decay.  It also required two charged particle tracks in the
spectrometer; a positive one with high momentum that continues
along the beam direction and goes into the calorimeter hole, and a
negative one that hits the calorimeter.  In addition, to avoid
accidental activity, trigger vetoes required that at least 18 GeV of energy be
deposited in the calorimeter, that no photon or charged particle 
escaped the fiducial volume of the detector and that there was no
hadronic activity outside the beam region.

Additional selection criteria for both signal and normalizing mode
were implemented offline in a non-blind analysis based on studying simulation
data.  The events were
required to have a positive particle track with 150 to 450 GeV/$c$
momentum and a negative particle track with 10 to 50 GeV/$c$ momentum.
The energy of each neutral cluster was required to be above 3 GeV,
separated by more than 15 cm from each other and more than 10 cm from
the track hit in the calorimeter. Using the energy and position of
these neutral clusters and assuming they proceeded from a $\pi^0$
decay, we reconstructed the decay vertex of the $\Sigma^+$ along the
positive particle (proton) track and reconstructed the $\Sigma^+$
four-momentum. The primary $\Xi^0$ decay vertex was then defined as
the point of closest approach of the extrapolated $\Sigma^+$ path and
the negative particle track ($\mu^-$ or $e^-$), also allowing us to
reconstruct a visible four-momentum for the $\Xi^0$ (missing the
neutrino).

All vertices were required to fall within the decay region (95 to 158
m), and the primary $\Xi^0$ vertex was required to lie within the
neutral beam fiducial volume and be upstream of the $\Sigma^+$
vertex. The $\Xi^0$ muon decay was distinguished from the $\Xi^0$ beta
decay by using the response of the calorimeter and the muon
identification systems.  The $\mu^-$ hit the calorimeter, depositing
minimum ionizing energy ($<$0.8 GeV), and was detected in the muon
system by 3 or more hit paddles (at least one in each of $\mu2$,
$\mu3x$ and $\mu3y$ planes). In addition the projected segment of the
negative particle track had to match the hits in the muon system
within 20~cm in the $\mu$2 plane and within 25~cm in the $\mu$3 planes
(allowing for known effects from multiple scattering).  The $e^-$ in
the normalizing $\Xi^0$ beta decay was identified by requiring that
its energy deposited in the calorimeter did not differ by more than
6\% from its measured momentum.  Also, to reject $K_L$ backgrounds the
momentum ratio between positive and negative particle tracks was
required to be greater than 4.5.  Further requirements were imposed on
both reconstructed $\Sigma^+\mu^-$ and $\Sigma^+e^-$ momentum (160 to
500 GeV/$c$) and the distance between the target and the primary
$\Xi^0$ vertex ($<$12 $\Xi^0$ lifetimes).

We also implemented some selection cuts considering specific
background decays. Possible backgrounds for $\Xi^0$ muon decay are:
$\ktpi$ $(K3\pi)$, $\kmu4$ $(K\mu4)$, $\lamppi$ and $\lammu$ decays
plus accidental $\gamma$s (accidental clusters that pass photon
identification cuts), $\caslpi$ with either $\lammu$ or $\lamppi$ as
subsequent decays, $\cassg$ followed by $\siglam$ and $\lamppi$,
$\caslg$ followed by $\lamppi$ plus an accidental $\gamma$, and
$\casbet$ plus an accidental $\mu^-$ and missing $e^-$. Here we are
assuming the $\pi^-$ decays into $\mu^-\bar{\nu}_\mu$ or punches
through the muon system faking a muon.

To reject $\caslpi$ background candidates, selection cuts were
implemented on the reconstructed $p\pi^-$ mass ($<$1.110 GeV/$c^2$)
and $\Lambda\pi^0$ mass ($>$1.330 GeV/$c^2$). The $\ktpi$ background
was rejected using a cut on the reconstructed $\pi^+\pi^-\pi^0$ mass
($>$0.600 GeV/$c^2$).  Furthermore in all background modes listed above
(except the $\Xi^0$ beta decay which is easy to distinguish from the
signal when the $e^-$ is detected) the two charged particles originate 
from a single vertex.  We found that a selection cut on the square of the
total transverse momentum ($>$0.018~(GeV/$c$)$^2$) of the two charged
particles relative to a line from the target to this vertex could provide
additional discrimination of these backgrounds from the signal mode.

\begin{figure}
\includegraphics[bb=0 25 534 325,width=8.4cm]{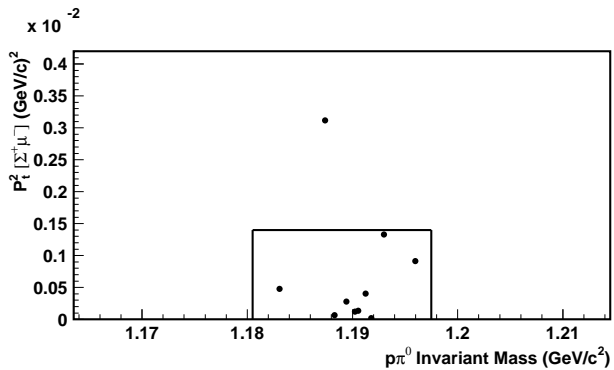}
\caption{\label{fig:events} Reconstructed $p\pi^0$ invariant mass and
$p^2_t$ of the $\Sigma^+\mu^-$ for events passing the $\Xi^0$ muon
decay selection criteria.}
\end{figure}

The distribution of events that survived all selection cuts is shown
in a plot of $p_t^2$, the square transverse momentum relative to the beam,
of the $\Sigma^+\mu^-$ versus the invariant mass of the p$\pi^0$ system
(Figure~\ref{fig:events}). The box is defined by the Monte Carlo
simulation to accept 90\% of the events. There are 9 signal events
inside the box (dots) clustered around 1.189 GeV/$c^2$ and 1 event
outside.

\begin{table}[b]
\caption{\label{tab:cuts}The effect on the number of signal events as a
  function of $\pm 1 \sigma$ variations in analysis cuts.  
  The cuts are on invariant mass
  combinations, as indicated in column 1.}
\begin{ruledtabular}
\begin{tabular}{lllll}
invariant         & cut variation&           & MC         & Data     \\
mass              &(GeV/$c^2$)   &           & expected   & observed \\ 
                  &              &           & events     & events   \\ \hline
                  & looser     & tighter     &  \\ \hline
p$\pi^-$          & $<$1.120   & $<$1.100    & 9.8 to 6.8 & 9 to 6   \\
$\Lambda\pi^0 $   & $>$1.322   & $>$1.338    & 9.6 to 5.4 & 9 to 5   \\
$\Sigma^+\mu^-$   & 1.285-1.323& 1.295-1.313 & 9.0 to 8.3 & 9 to 8 \\
$\pi^+\pi^-\pi^0$ & $>$0.540   & $>$0.660    & 9.0 to 8.7 & 9   \\
\end{tabular}
\end{ruledtabular}
\end{table}

In order to test the robustness of the selection of our signal events
and the Monte Carlo simulation we varied the cuts on reconstructed
masses by one sigma, observing how this affected the number of events
in the signal box for both data and Monte Carlo events (Table
\ref{tab:cuts}). We found that in all cases the number of observed
signal events followed the prediction of the Monte Carlo, also
implying negligible background among the signal events.  Moreover, the
likelihood of the two-dimensional distribution of the 9 found events
was compared to a toy Monte Carlo of 1,500 equivalent 9-event
experiments. We found that 31\% of the MC experiments had a lower
likelihood than the data, indicating a high degree of correspondence between
the distribution of these events and the expectation.

\begin{figure}
\includegraphics[bb=0 25 534 325,width=8.4cm]{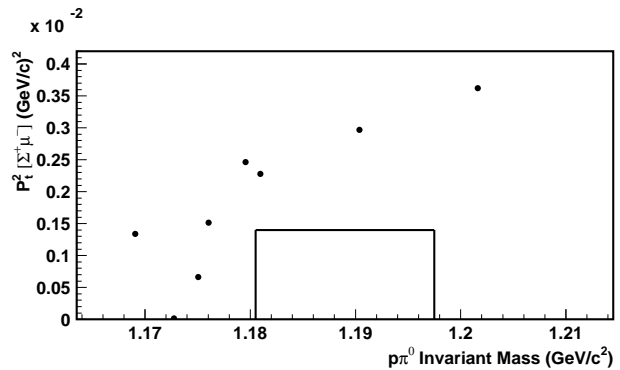}
\caption{\label{fig:cassg} Reconstructed $p\pi^0$ invariant mass and
$p^2_t$ of the $\Sigma^+\mu^-$ for MC $\cassg$ events passing the
$\Xi^0$ muon decay selection criteria.  Ten times the
measured number of $\Xi^0$ decays were generated.}
\end{figure}

Two techniques were used to investigate the background decay
modes. First, we used Monte Carlo to simulate several modes by
generating at least ten times the expected number of events. Second,
we employed a wrong-sign charge analysis using data to measure the
level of charge symmetric $\ktpi$ decays as background, since it would
be not feasible to simulate them by Monte Carlo due to the enormous
number of required events.  The wrong-sign analysis consisted of
selecting events having a negative high-momentum particle in the beam
satisfying the proton selection criteria and a positive low-momentum
particle in the calorimeter satisfying the muon identification
criteria.  This technique took advantage of the symmetry between
$\pi^+$ and $\pi^-$ in $\ktpi$ decay and of the suppression of
anti-hyperon production at this energy; the $\overline{\Xi}^0$
production rate being an order of magnitude smaller than the $\Xi^0$
rate~\cite{monnier}.  There were no background events seen from the
Monte Carlo or from the wrong-sign analysis inside the signal box
after applying all selection criteria. Using Monte Carlo we estimate
aproximately one background event outside the signal box from $\cassg$
followed by $\siglam$ and $\lamppi$ and no background events from all
the other decay modes investigated. This estimate for background is
shown in Figure~\ref{fig:cassg}, which shows 8 events surviving the selection
cuts outside the signal region. This is from a sample of simulated
$\Xi^0$ decays that has ten times the expected number of events in the
real experiment.

\begin{figure}
\includegraphics[bb=0 20 534 300,width=8.4cm]{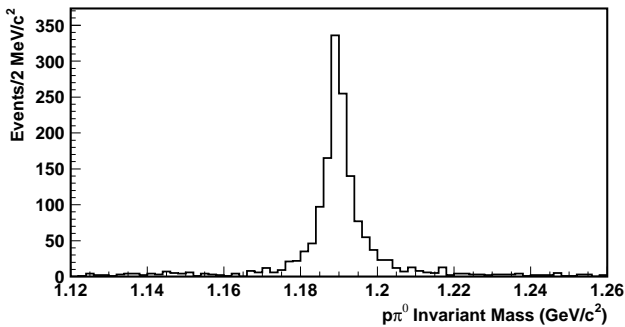}
\caption{\label{fig:casbet} Reconstructed $p\pi^0$ invariant mass
  distribution for 1999 data events passing the $\casbet$ selection
  criteria.}
\end{figure}

\begin{figure}
\includegraphics[bb=0 20 534 330,width=8.4cm]{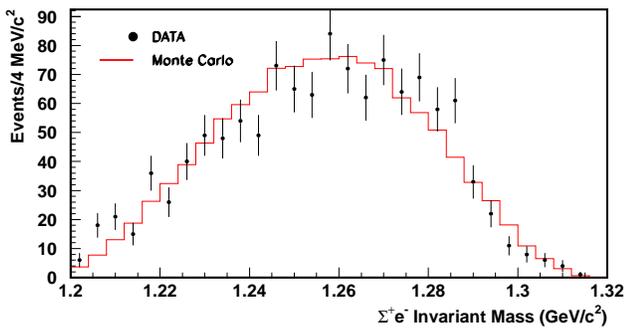}
\caption{\label{fig:mcasbet} Data-MC comparison of $\Sigma^+e^-$
  invariant mass distribution for events passing the $\Xi^0$ beta
  decay selection criteria. Data are shown in dots; MC in histogram.}
\end{figure}

The $\Xi^0$ beta decay, used for normalization, yielded 1139 events,
after subtracting 54 background events within a window of 17 MeV/$c^2$
around the central value of the p$\pi^0$ invariant mass distribution
(Figure~\ref{fig:casbet}). A comparison between data and Monte Carlo
of the selected events (before the background subtraction) was
implemented for the $\Sigma^+e^-$ invariant mass which does not
reconstruct the $\Xi^0$ mass since we miss the neutrino energy
(Figure~\ref{fig:mcasbet}).  Monte Carlo studies gave the acceptances
of both signal and normalizing modes: A($\casmu$) = 1.36\% and
A($\casbet$) = 3.01\%.  The acceptances included losses due to
detector geometry, trigger efficiencies, event reconstruction and
particle identification.

We find the ratio of decay rates $\Gamma(\casmu)/\Gamma(\casbet)$ to
be $\left(1.8~^{+0.7}_{-0.5}~(stat.) \pm0.2~(syst.)\right) \times
10^{-2}$. The statistical uncertainties were determined at a 68\%
confidence level with the Feldman and Cousins method \cite{feldman}.
The systematic error is the quadratic sum of $\pm 0.5
\times 10^{-3}$ from the uncertainty in the number of normalizing mode
decays, $\pm 0.6 \times 10^{-3}$ from the uncertainty in the 
ratio of acceptances
and $\pm 1.4 \times 10^{-3}$ due to an 8\% muon identification
uncertainty.  This ratio of decay rates is consistent with the
calulation assuming lepton flavor symmetry.

Using the published value of the normalizing $\casbet$ branching
fraction, $(2.7 \pm 0.4) \times 10^{-4}$ \cite{pdg}, we find the
branching fraction of the $\Xi^0$ muon semi-leptonic decay to be
$\left(4.7~^{+2.0}_{-1.4}~(stat.) \pm0.8~(syst.)\right) \times 10^{-6}$ at the
68\% C.L., where the systematic error also includes the contribution
due to the uncertainty of the $\Xi^0$ beta decay branching fraction.

In summary, we have observed the $\Xi^0$ muon semi-leptonic decay and
measured its decay rate using the $\Xi^0$ beta decay as normalizing
mode. This observation agrees with the SU(3) flavor quark model
prediction~\cite{garcia} and the assumption of lepton flavor symmetry.

\begin{acknowledgments}
We gratefully acknowledge the support and effort of the Fermilab staff
and the technical staffs of the participating institutions for their
vital contributions. This work was supported in part by the
U.S. Department of Energy, The National Science Foundation, The
Ministry of Education and Science of Japan, Funda\c c\~ao de Amparo
\`a Pesquisa do Estado de S\~ao Paulo (FAPESP), Conselho Nacional de
Desenvolvimento Cient\'{\i}fico e Tecnol\'ogico - Brazil (CNPq) and
The Ministry of Education of Brazil (CAPES).  In addition,
R.A.G. acknowledges the valuable discussions with P.\,S. Cooper.
\end{acknowledgments}

\end{document}